\documentclass{article}
\usepackage{graphicx}	
\usepackage{amsmath}
\usepackage{amssymb}
\usepackage{mathrsfs}
\usepackage{float}
\usepackage{color,soul}
\usepackage[title]{appendix}
\usepackage[a4paper, total={6in, 9in}]{geometry}
\usepackage{authblk}
\usepackage{multicol}
\usepackage{comment}
\usepackage[font={footnotesize}]{caption}
\usepackage{nameref}
\usepackage{textcase}

\newcommand{\half}{\frac{1}{2}}
\newcommand{\abs}[1]{\left| #1 \right|}
\newcommand{\ket}[1]{\left| #1 \right>}
\newcommand{\bra}[1]{\left< #1 \right|}
\newcommand{\expect}[1]{\left< #1 \right>}
\newcommand{\round}[1]{\left( #1 \right)}

\def\ii{{\rm i}}  \def\ee{{\rm e}}
\def\rb{{\bf r}}  \def\Rb{{\bf R}}    \def\vb{{\bf v}}
        
\def\kb{{\bf k}}    
    
     \def\qb{{\bf q}}     

\begin{document}

\title{Optical coherence transfer mediated by free electrons}
\author[1,2]{Ofer Kfir \thanks{ofer.kfir@mpibpc.mpg.de}}
\author[3]{Valerio Di Giulio}
\author[3,4]{F. Javier Garc\'{i}a de Abajo}
\author[1,2]{Claus Ropers}

\affil[1]{University of Göttingen, IV. Physical Institute, Göttingen,
Germany }
\affil[2]{Max Planck Institute for Biophysical Chemistry (MPIBPC), Göttingen, Germany }
\affil[3]{ICFO-Institut de Ciencies Fotoniques, The Barcelona Institute of Science and Technology, 08860 Castelldefels, Barcelona, Spain}
\affil[4]{ICREA-Instituci\'{o} Catalana de Recerca i Estudis Avançats, Passeig Lluís Companys 23, 08010 Barcelona, Spain}

\date{\today}
\maketitle

\begin{abstract}
We investigate theoretically the quantum-coherence properties of the cathodoluminescence (CL) emission produced by a temporally modulated electron beam. Specifically, we consider the quantum-optical correlations of CL from electrons that are previously shaped by a laser field. The main prediction here is the presence of phase correlations between the emitted CL field and the electron-modulating laser, even though the emission intensity and spectral profile are independent of the electron state. In addition, the coherence of the CL field extends to harmonics of the laser frequency. Since electron beams can be focused to below one Angstrom, their ability to transfer optical coherence could enable ultra precise excitation, manipulation, and spectroscopy of nanoscale quantum systems.
\end{abstract}




Inelastic electron scattering constitutes the basis of various powerful spectroscopy and spectrally selective imaging techniques \cite{Polman_Kociak_JGdA_NatMat_2019}. Cathodoluminescence (CL) and electron energy-loss spectroscopy (EELS) are related approaches for harnessing the spectral density of spontaneous interaction processes \cite{Reimer_Kohl_TEM_book_2008}, where their stochastic nature causes a loss of coherence. In CL, there is no external field other than that provided by the electron, so there is no reference to exhibit coherence with. In EELS, the incident and inelastically scattered electron states lose their coherence through the random phase associated with the excitation. 
A stimulated counterpart for EELS interactions has been established in the form of photon-induced near-field electron microscopy (PINEM) \cite{Barwick_PINEM_2009,Park_NJP_PINEM_2010, JGdA_NanoLett_2010}, or electron-energy gain spectroscopy (EEGS) \cite{H99,EEGS_Javier_Kociak_NJP_2008}. In these techniques, the external excitation of a particular mode increases its interaction probability and thus selectively enhances the sensitivity for probing or imaging \cite{Piazza_EEGS_antenna_NatCom_2015,DAS_Kociak_EEGS_Ultramic_2019}. The coherence of the electron-energy states is evident in a transverse \cite{vanacore_transverse_PINEM_NatComm_2018, vanacore_PINEM_vortex_NatMat_2019, feist_transverse_PINEM_2020} or  longitudinal structuring of the electron beam as attosecond pulses \cite{ Feist_Nature_2015, Priebe_atto_SQUIRRELS_NatPhot_2017, Morimoto_atto_NatPhys_2018, Kozak_Hommelhoff_atto_PRL_2018}. Recent work \cite{Gover_and_Yariv_2LS_Rabi_after_PINEM_PRL2020} has used semi-classical arguments to conclude a dependence of the excitation of two-level systems on the electron wave function, although such dependence disappears in a full quantum treatment of the system. A quantum description of CL properties should adhere to our current understanding of CL, for which a point-particle description of the electron is sufficient. By addressing the quantum nature of the electron, one can ask how and to what extent would properties of CL, such as its intensity, coherence characteristics, and radiation pattern \cite{Coenen_Polman_APL_2011,Osorio_Polman_ACSPhot_2016}, be affected by the incident state of the electron? \\

Here, we directly address these questions in a rigorous theoretical framework, making predictions for the quantum state of radiation produced by phase- and density-modulated electron states. Establishing that an electron beam can coherently stimulate optical excitations, we introduce the notion of “electron-mediated coherence transfer”. Specifically, our results show that temporal shaping of an electron beam has profound and measurable consequences for inelastic electron-light scattering. In the single-electron limit, we demonstrate that the coherence properties and the quantum optical correlations strongly depend on the details of the electron state, while the CL intensity and spectral profile remains unaffected by the electronic wave function.
In particular, we show that CL from PINEM-modulated electrons can exhibit mutual coherence with a replica of the PINEM-driving optical field, or with its harmonics. We propose interferometric measurements for the extraction of phase information in CL by heterodyne tomography of the radiation quantum state. The results are readily applicable not only to generated radiation, such as CL, but also to non-radiative excitations (e.g., dark polaritons \cite{Chu_JGdA_Dark_plasmon2009}). This concept defines a way of transferring optical polarization carried by electrons with sub-nm precision, which has a potential for accessing and manipulating individual quantum systems.

\begin{figure}[H]
    \centering
    \includegraphics[width=11cm]{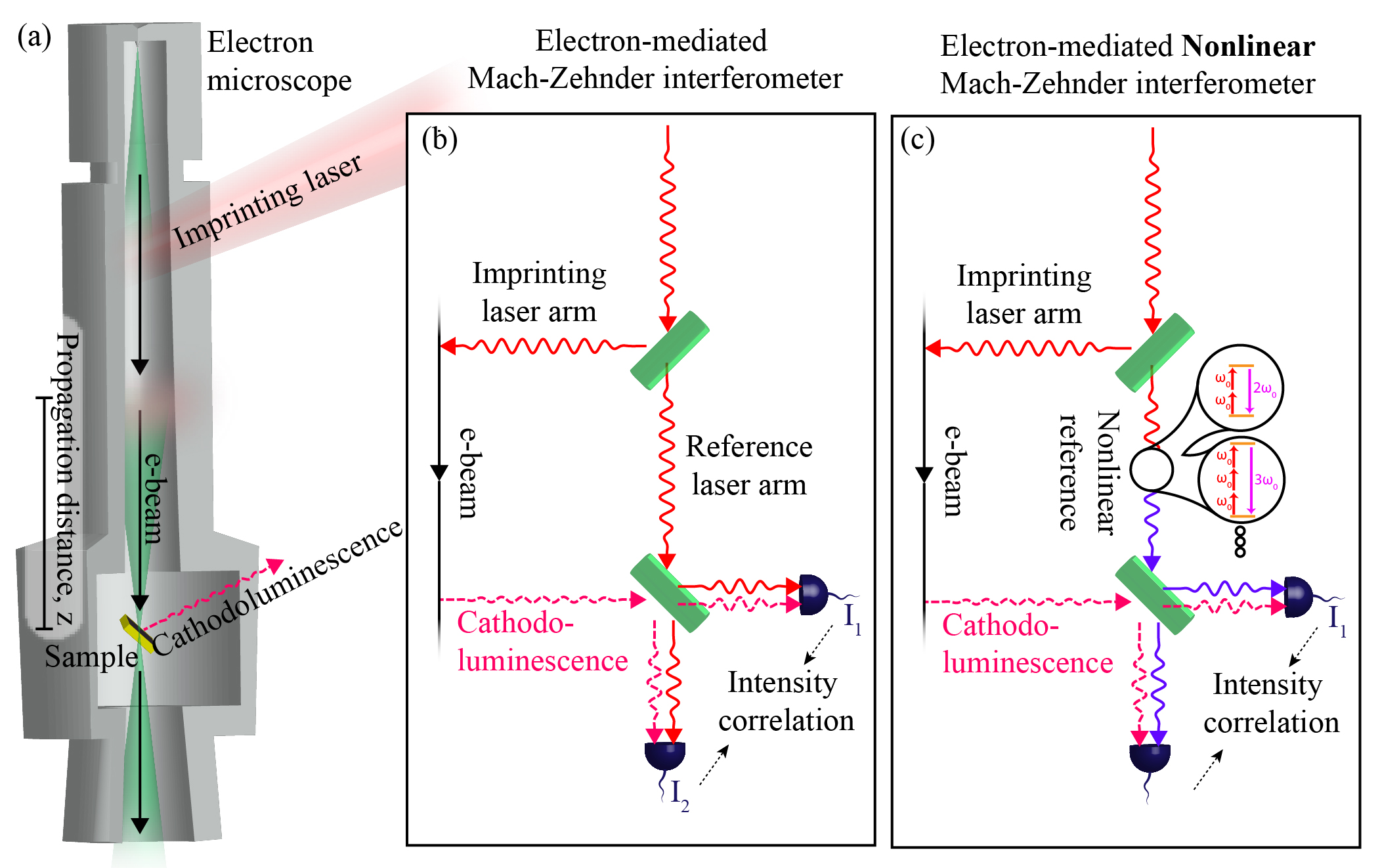}
    \caption{Linear and nonlinear Mach-Zehnder interferometer incorporating a free-electron beam. (a) Proposed experimental concept based on an electron beam (illustrated in green) in a transmission electron microscope (TEM). A laser field imprints optical-phase information on the electron, which after propagation can transfer it back to the radiation via cathodoluminescence (CL) emission, typically near the sample section of the microscope. (b) Scheme for a Mach-Zehnder interferometer with a reference laser and optical coherence carried by a free-electron beam. Intensity correlations between the two interferometer ports, measured as $I_1$ and $I_2$, are used to retrieve information on the electron arm of the interferometer, as well as on the sample. (c) Nonlinear Mach-Zehnder interferometry can reveal information on higher-order components of the electron state and of the CL by interfering with harmonic frequencies of the reference field.}
    \label{fig:TEM_MachZehnder}
\end{figure}

Figure  \ref{fig:TEM_MachZehnder} represents a conceptual system to investigate coherent CL, which can be implemented within an electron microscope. Optical phase information from a PINEM-driving laser field is imprinted on and carried by the electron over a distance $z$, resulting in a coherent CL emission by interaction with an out-coupling sample system. Such electron transfer of optical coherence can be detected by an interferometric setting that targets either the linear or the nonlinear response of free electrons (Figs.  \ref{fig:TEM_MachZehnder}b and  \ref{fig:TEM_MachZehnder}c, respectively).\\

We first consider comb-like electron energy superposition states combined with the radiation vacuum $\left|0\right>$,
    \begin{equation}
        \left| \psi_{in} \right\rangle =      \left| 0  \right\rangle \otimes \sum^\infty_{j=-\infty} c_j \left| E_j\right\rangle   = \sum^\infty_{j=-\infty} c_j \left| E_j , 0\right\rangle 
        \label{eq:e_ladder_state} 
    ,\end{equation}
where $c_j$ are the complex probability amplitudes for electron states with energy $E_j$, and the index $j$ runs over electron-energy levels. The coefficients $c_j$ are normalized as $\sum_j \left| c_j  \right|^2=1$, and their phases vary with electron propagation in vacuum according to the free-particle dispersion. The quantum-optical and coherence properties of the CL from an electron state with a temporally modulated density depend on the interaction of the electron with an emitter. For simplicity, we consider a single non-degenerate optical band into which CL is emitted. General expressions for the quantum properties of the CL are derived in the Appendix, including the radiation continuum, as well as states with definite momentum, energy, and polarization. The photon frequency represents a good quantum number within the interaction bandwidth because it is a single-valued function of the longitudinal momentum, while transverse deflections can be neglected under the nonrecoil approximation \cite{Javier_rev_mod_ph_2010}. The quantum-optical properties of the CL emission can be described by the scattering operator, $\hat{S}$, which under the above conditions has the form of a displacement operator,
\begin{equation}  
\hat{S} = \exp\left[ \int_0^{\infty} d\omega  \left( g_\omega \hat{b}_\omega \hat{a}_\omega^\dagger  -  g_\omega^*\hat{b}_\omega^\dagger \hat{a}_\omega  \right)   \right]
\label{eq:S_operator},
\end{equation}
as derived for classical fields \cite{Feist_Nature_2015}. For every frequency, the scattering operator allows for annihilation and creation of a photon, marked by the  operators $\hat{a}_\omega$ and $ \hat{a}_\omega^\dagger$, where  energy is conserved by the corresponding electron-energy ladder operators $ \hat{b}_\omega^\dagger$ and $\hat{b}_\omega$, respectively. Here, $g_\omega$ accounts for the electron-photon coupling at the angular frequency $\omega$, where the photon spectral density of the CL is given by $\left|g_\omega\right|^2$. Similarly, $\left|g_\omega\right|^2$ would be the EELS spectral density in the absence of competing loss mechanisms (e.g., bulk plasmons). The phase of $g_\omega$ is arbitrary, and $g_\omega$ can be chosen as a non-negative real-valued spectral function. However, in examples such as the radiation into normal modes of a fiber, it is convenient to impose a flat spectral phase on the photonic modes and place the spectral degree of freedom as a complex coupling function. A rigorous derivation of Eq. \eqref{eq:S_operator} as well as the conditions for which it applies are given in Appendix \ref{Appendix:Scatt_operator_derivation}. 

The final quantum state is $\left|\psi_{f}\right> =\hat{S}\left|\psi_{in}\right>$, which can be calculated for an arbitrary coupling strength, is used to obtain the properties of the CL emission (see Appendix \ref{Appendix:strong_e_ph_coupling}). One can acquire some intuition from the first-order approximation in the weak electron-photon coupling regime, $\left|g_\omega\right|^2\ll1$, where the final state is  
    \begin{equation}\label{eq:psi_final} 
        \left|  \psi_{f}    \right\rangle \approx \sum_j c_j \left[ \left|  E_j,0    \right\rangle  + \int d\omega g_\omega \left| E_j -\hbar\omega, 1_\omega     \right\rangle       \right].
    \end{equation} 
The weak interaction has a small probability amplitude to create a photon with angular frequency $\omega$, represented by state $\left|1_\omega\right>$, accompanied by a corresponding electron-energy loss. Notably, the CL intensity is unaffected by the specific electron superposition state,
\begin{align}
\label{eq:n_w}
  \left\langle    \hat{n}_\omega  \right\rangle &= \left\langle \hat{a}_\omega^\dagger \hat{a}_\omega  \right\rangle 
  =
  \sum_{j=-\infty}^\infty \left|   c_j   \right|^2 \left|    g_\omega  \right|^2 = \left|   g_\omega   \right|^2 ,  
\end{align}
(see derivation for continuous spectrum in Appendix \ref{Appendix:CL_photon_number}). The expectation value for an operator is given with respect to the final electron-photon state, $\left\langle \hat{O} \right\rangle \equiv \left\langle \psi_{f} \left|   \hat{O}\right| \psi_{f}   \right\rangle$. Equation \eqref{eq:n_w} complies with the current understanding of CL, and thus, provides for a solid scientific basis for the predictions in this work. In contrast to the wave-function-independent photon emission probabilities, the expectation value for the electric field carries information on the electron temporal structure. The physical electric field at a particular frequency $\omega$ and time $t=0$ is $\vec{E}_{\omega }=\left< \hat{ \vec{E}}_{\omega } \right>$, can be represented as a sum of two complex components, $\hat{\vec{E}}_{\omega }=\hat{\vec{E}}_{\omega }^{(+)}+\hat{\vec{E}}_{\omega }^{(-)}$, with $\hat{\vec{E}}_{\omega }^{(-)}=(\hat{\vec{E}}_{\omega }^{(+)})^\dagger$. The field is proportional to the ladder operator, and we have 
  \begin{equation}
  \left<    \hat{E}_{\omega }^{(+)}  \right> 
  \propto \left<  \hat{a}_{\omega }   \right>  
.\end{equation}
The main reason to represent the field with the ladder operator, $\hat{a}_\omega \equiv \hat{a}_{\omega,(t=0)}$, is that $\hat{a}_\omega$ relates directly to the photon statistics (e.g., shot noise in an interferometer) and it is proportional to the field. This procedure is legitimate if the effects of the spatial distribution and the polarization of the field can be traced out, as for example, in CL into a single-mode fiber (see the exact expression in Appendix \ref{Appendix:Mean_electric_field_Valerio}). Evaluating the photonic ladder operator with respect to $\left|\psi_{f}\right>$ in Eq. \eqref{eq:psi_final}, we find
    \begin{align} \label{eq:a_as_cj_cj_gw}
        \left<\hat{a}_\omega\right> 
        &= \sum^\infty_{j,j'=-\infty} c_j^*c_{j'} g_\omega \left<E_j,0 \left| E_{j'}-\hbar\omega,0\right.\right>. 
    \end{align}
We assume that the PINEM-driven state is a comb separated by the photon energy, $\hbar\omega_0$, where the electron spectral density distribution is much narrower than the separation of the levels. Thus, one can use discrete level indices \cite{OPticaDIgiulio} and write $\left<E_j \left| E_{j'}-\hbar\omega\right.\right>
 =  \delta_{E_j,E_{j'}-\hbar\omega}
 =  \delta_{j',j+n}$, where $n$ is the energy exchange in terms of a harmonic of the fundamental ladder separation, $n=\frac{\omega}{\omega_0}$.  
  Substituting the Kronecker $ \delta $ into Eq. \eqref{eq:a_as_cj_cj_gw} yields a simple expression for the field,
\begin{equation}
   \left<\hat{a}_\omega\right> =
   \begin{cases}
   g_{n\omega_0} \sum^\infty_{j=-\infty} c_j^*c_{j+n} & \omega=n\omega_0\\
   0 & \text{otherwise}
   \end{cases}
   , \,n \in \mathbb{Z}
   \label{eq:a_vs_omega}
   .\end{equation}
Since $c_j$ are the amplitudes of the energy states $E_j$, they are proportional to the Fourier coefficients of the electron wave function, that is, $c_j \equiv \expect{E_j|\psi} \propto \int \psi(t) e^{\ii j \omega_0 t} dt $, which can be used to simplify the CL field 
\begin{equation}  
\left< \hat{E}_\omega^{(+)} \right>  \propto  \left< \hat{a}_{\omega} \right> = g_{\omega} \mathcal{FT} \left[  \left| \psi (t) \right|^2 \right]_{(\omega)} 
\label{eq:E_g_nw_FT_psi_sq}
.\end{equation}
Incidentally, the temporal electron-probability amplitude, $\psi(t)$ can be represented spatially along the propagation axis $\tilde{\psi}(z)=\psi(t=z/v_e)$ using the electron group velocity $v_e$. Equation \eqref{eq:E_g_nw_FT_psi_sq} is a central result of this paper, representing a general property of CL from a structured electron state (see detailed calculation in Appendix \ref{Appendix:Mean_electric_field_Valerio}). We emphasize that only the expectation value of the field follows the electron density, whereas the mean photon number, $\left\langle \hat{n}_\omega \right\rangle$, is unaffected by the electron temporal structure. Additionally, higher-order correlations are more intricate (see derivation in Appendix \ref{Appendix:higher_order_correlations}),
\begin{equation}
\begin{split}
\left<\left(\hat{a}-\left<\hat{a}\right>\right)^N\right> = 
g_\omega^N \sum_k\binom{N}{k}\mathcal{FT}\left[\left|\psi(t)\right|^2\right]_{\left(k\omega\right)} 
\cdot \left(-\mathcal{FT}\left[\left|\psi(t)\right|^2\right]_{\omega}\right)^{N-k}
,\end{split}
\end{equation}
where $\binom{N}{k}$ are Newton's binomial coefficients. 
To isolate the coherence properties of the emitted CL, we define the degree of coherence (${\rm DOC}$) as the power associated with the phase-carrying field compared to the overall energy \cite{Mandel_and_Wolf}. More precisely,  
\begin{align}
    {\rm DOC}(\omega)  &= \frac{ \left\langle \hat{E}_\omega^{(-)}  \right\rangle \left\langle \hat{E}_\omega^{(+)}  \right\rangle }{\left\langle  \hat{E}_\omega^{(-)} \hat{E}_\omega^{(+)} \right\rangle }
  = \frac{\left\langle \hat{a}_\omega^\dagger \right\rangle  \left\langle \hat{a}_\omega \right\rangle}{\left\langle \hat{a}_\omega^\dagger \hat{a}_\omega \right\rangle} \label{eq:defining_DOC} 
    = \left| \mathcal{FT} \left[  \left| \psi (t) \right|^2 \right]_{(\omega)} \right|^2  
.\end{align}
Obviously, for a coherent state of light \cite{Glauber_coherent_1963}, the degree of coherence is unity. Furthermore, the above expression remains unchanged for a continuous electron spectrum as expressed in Appendix \ref{Appendix:DOC_Continuum_states_Valerio}. \\

Let us further explore the predictions of Eq. \eqref{eq:defining_DOC}. 
 First, the degree of coherence depends purely on the properties of the luminescing electron, and not on the material or the geometry of the electron-light coupler. Second, we have ${\rm DOC}(\omega=0)=1$, since the electron wave function is normalized. Third, the CL \textit{field} is proportional to $\sqrt{{\rm DOC}(\omega})$, and thus, to the electron \textit{probability density}. The coherent fraction of the CL power is proportional to the square of the temporal electron density. Therefore, dense electron distributions are preferred, as in the form of short electron pulses (e.g., full width at half maximum (FWHM) of 200 fs, as experimentaly shown in Ref. \cite{Feist_UTEM_Ultramicroscopy_2017}). We note that the coherent CL field emitted from such a pulsed electron must be pulsed as well, with an equal duration, thus, the FWHM of the optical \textit{intensity} is shorter by a factor of $\sqrt{2}$.\\ 

\begin{figure}[H]
    \centering
    
     \includegraphics[width=11cm]{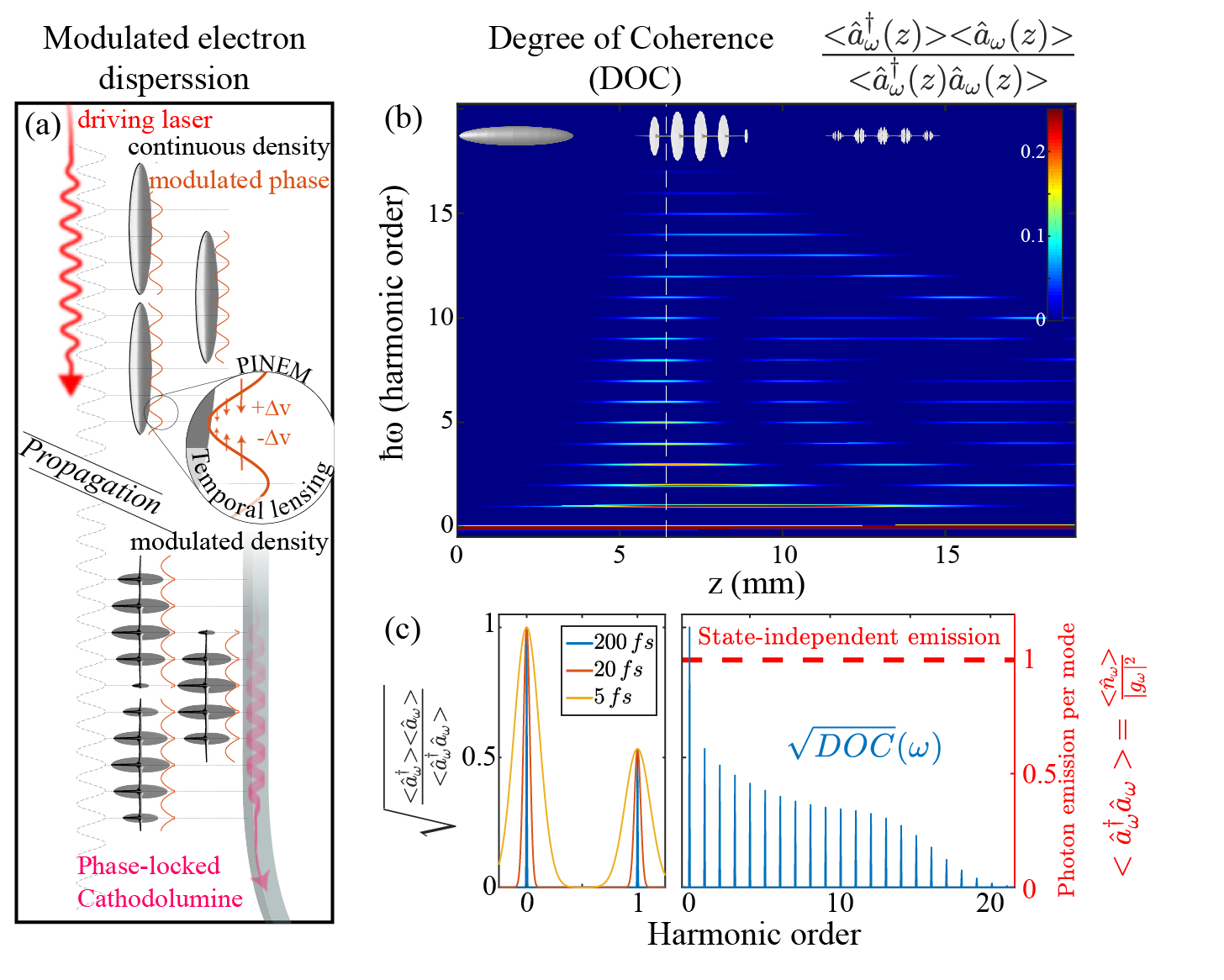}
    
    \caption{Coherent properties of cathodoluminescence emission by shaped electrons. (a) (top) Laser-driven PINEM imprints phase oscillations on an electron wavepacket, regardless of the exact arrival timing of the electron. The electron-phase oscillations are equivalent to a temporal lens, slowing and accelerating parts of the wavepacket periodically. (bottom) After propagation, anharmonic density and phase modulations evolve in the electron, leading to coherent CL (e.g., in a waveguide). (b) Calculated evolution of the degree of coherence (${\rm DOC}$) following PINEM  as a function of the propagation distance $z$. The ${\rm DOC}$ is nonzero only for harmonics of the PINEM-driving laser. (c) The emitted field is proportional to $\sqrt{\rm DOC}$, and its width is inversely proportional to the pulse duration of the pre-structured electron density (see legend for FWHM electron-pulse durations). While the ${\rm DOC}$ and its square root comprise a harmonic frequency comb that varies with $z$, the emission rate of CL photons (dashed red line) remains unaffected by the temporal structure and only depends on the coupling amplitude, $g_\omega$, between the electron and the optical modes. }
    \label{fig:DOC}
\end{figure}

 Figure \ref{fig:DOC} presents a sinusoidal modulation of the electron through PINEM, and the resulting properties of the ${\rm DOC}$ for the emitted radiation. In particular, Fig. \ref{fig:DOC}a illustrates a PINEM-driving laser field that imprints an oscillatory phase (orange) on the electron, without an immediate change on the electron density (gray). The electron-phase curvature represents a varying velocity, $\Delta v$, which acts as temporal lensing that modulates the electron probability density after propagation. Cathodoluminescence (e.g., into an aligned waveguide) should be locked to the phase of the modulated electron, and therefore, to the phase of the driving field. The electron arrival time does not affect the locking to the phase of the PINEM-driving laser, as illustrated by the few electron replicas sketched in Fig. \ref{fig:DOC}a. Figure \ref{fig:DOC}b shows the ${\rm DOC}$ for CL from PINEM-modulated electron, as a function of the propagation distance, $z$, to the CL emitter. For this numerical simulation, we used a standard electron beam acceleration voltage of 200 keV, a wavelength of 800 nm ($\hbar\omega_0=1.55 \,eV$) and an easily accessible PINEM parameter $\abs{\beta} = 4$, such that the phase at $z=0$ is given by $e^{-2 \abs{\beta} \ii \cos{\omega_0 t}}$. In particular, Fig. \ref{fig:DOC}b presents the sum $\sum_j \left(c_j(z) \right)^* c_{j+n}(z)$, which is nonzero only for harmonics of the modulating laser, as predicted by Eq. \eqref{eq:a_vs_omega}. Here, we take the initial amplitudes of the electron state as Bessel functions of the first kind \cite{Feist_Nature_2015,Kfir_e_ph_entanglement_PRL2019}, $c_j(z=0)=J_j(2|\beta|)$. The coefficients $c_j$ evolve in vacuum as $c_j(z)=c_j(0)e^{\ii k_z z}$, where $\hbar k_z$ is the electron momentum for energy $E_j$ and $\hbar$ is the reduced Planck constant (see Appendix \ref{Appendix:DOC_PINEM_explicit} for an explicit representation). Figure \ref{fig:DOC}c focuses on a particular propagation distance $z=6.43\,\rm mm$, where the CL spectrum is widest. The ${\rm DOC}$ at integer multiples of $\omega_0$ is fixed, aside form the $z$ dependence. The distribution in the vicinity of the harmonics is determined by the duration of the pre-structured electron, as plotted in Fig. \ref{fig:DOC}c for Gaussian electron pulses. Taking Fig. \ref{fig:DOC}c to two of its extremes, if the electron pulse duration is infinitely long, the ${\rm DOC}$ is nonzero only at an infinitesimal band for each harmonic. In contrast, if the electron is infinitely compressed to a point-like particle, the ${\rm DOC}$ of the various harmonics merge, and the CL is fully coherent. For a PINEM-driven modulation, the CL field (proportional to $\sqrt{\rm DOC}$) can reach nearly 50\% coherence for many harmonic orders (blue curve in Fig. \ref{fig:DOC}c). The CL emission rate, normalized to the coupling probability, is independent of frequency (red dashed curve), as expressed in Eq. \eqref{eq:n_w}.

\begin{figure}[H]
    \centering
    
    \includegraphics[width=11cm]{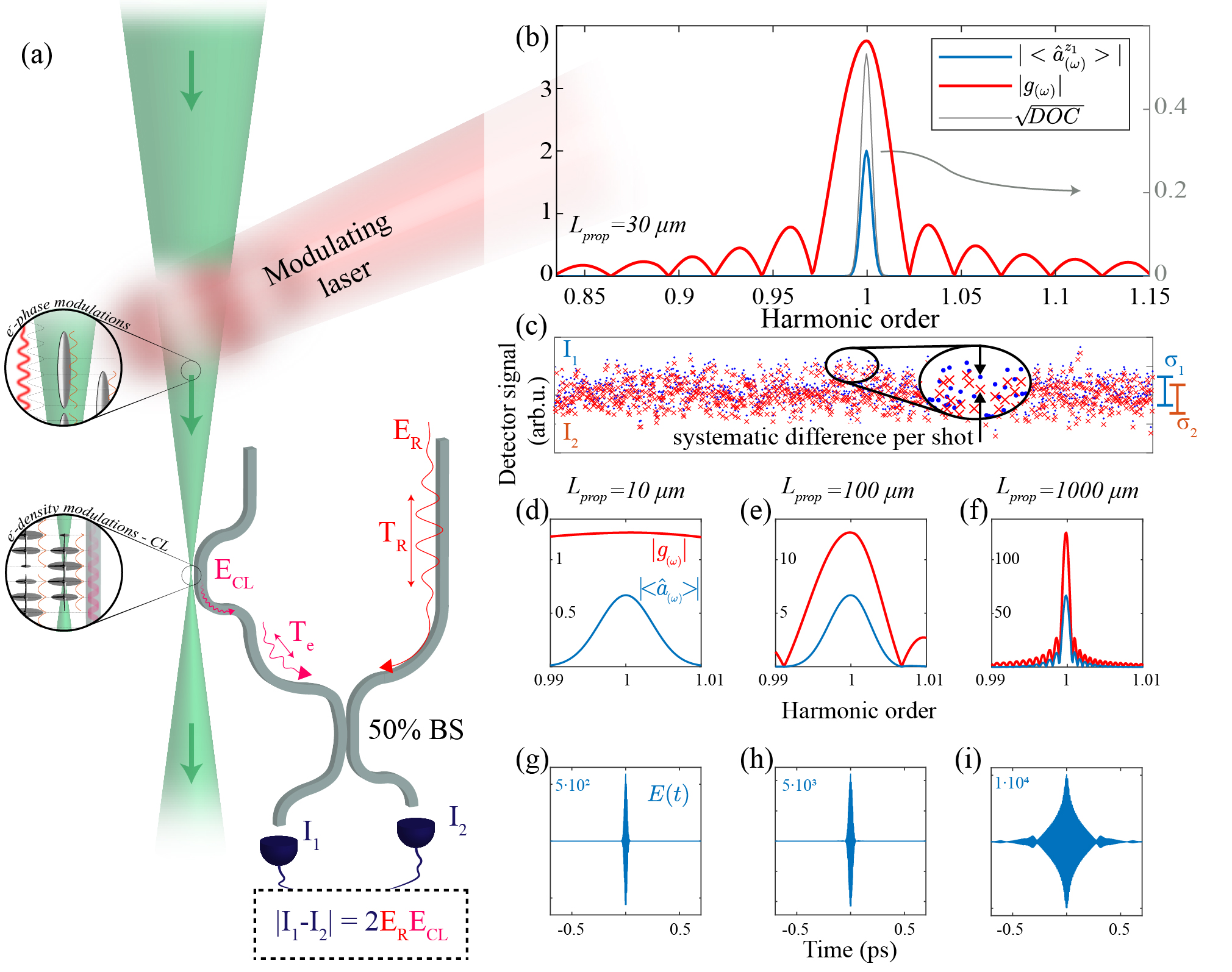}
    
    \caption{
    Correlations in the cathodoluminescence into a dielectric waveguide produced by modulated electrons. (a) Conceptual scheme of the experimental setting. A laser field structures a pulsed electron beam, which excites coherent radiation in a parallel-aligned waveguide. Mixing a replica of the driving laser field, $E_R$, with the CL field, $E_{CL}$, results in a difference between the signals recorded in the detectors at the output ports of the  interferometer, $I_1$ and $I_2$. (b) CL properties of emission into the waveguide for an interaction length of $30\, {\rm \mu m}$. The coherent CL amplitude (blue) is the product of the spectral coupling amplitude, $\abs{g_\omega}$ (red), and  $\sqrt{\rm DOC}$ (gray), where phase-matching between the electron and the guided modes limits the CL emission to a single harmonic peak. (c) The stochastic photon counts in the two detectors ($I_1$, blue; $I_2$, red) is correlated in the limit of a strong reference and ideal detectors. $\sigma_1$  and $\sigma_2$ denote the standard deviation (shot noise) at each respective detector. The systematic difference per shot is marked in the inset. Panels (d-i) show CL properties as a function of the propagation length near the fiber, $L_{prop}$: (d-f) show the spectral distribution of $\abs{g_\omega}$ and $\expect{\hat{a}_\omega}$, whereas (g-i) show the emitted CL-field pulse in the time domain, $E(t)$. The different columns show the CL properties for an electron propagation length $L_{prop} = 10\,\rm \mu m$ (d,g), $100\,\rm \mu m$ (e,h), and $1\,\rm mm$ (f,i) along the waveguide. When the coupling bandwidth is narrower than the width of the ${\rm DOC}$ (e.g., in panel (f)) the CL pulse is temporally distorted (panel (i)). Details of the coupling into a cylindrical waveguide are presented elsewhere \cite{Kfir_e_ph_entanglement_PRL2019}. The calculation for the stochastic signal assumes Poisson-distributed photon counts from an arbitrarily strong reference coherent state, with negligible additional noise from the CL field.  
    }
        \label{fig:CL_into_a_waveguide}
\end{figure}

In order to provide for a quantitative example, we consider CL into a parallel dielectric waveguide \cite{Bendana_Polman_JGdA_CL_fiber_NL_2010,Kfir_e_ph_entanglement_PRL2019} produced by a 200-fs-long, Gaussian electron pulse modulated by PINEM. The waveguide parameters are chosen such that the phase velocity of the optical mode at the frequency of the PINEM-driving laser equals the electron group velocity. For electrons accelerated to 200 keV (corresponding to 69\% of the speed of light in vacuum), we find that the coupling is most efficient for electrons passing near the surface of cylindrical silicon-nitride waveguides with a diameter of 345.8 nm. Figure \ref{fig:CL_into_a_waveguide} describes such an experiment. 
The CL field, $E_{CL}$, can be collected in a waveguide and mixed with a replica of the driving laser (red), $E_R$, using a 50/50 beamsplitter. Figure \ref{fig:CL_into_a_waveguide}b shows the coupling $\abs{g_\omega}$ alongside the emitted field $\left|  \left<\hat{a}_{ \omega} \right>\right|$ for electrons that propagate for $\rm 30\,\mu m$ near the waveguide surface. In such Mach-Zehnder interferometry between weak and strong inputs, the difference collected by two detectors, $I_1$ and $I_2$ (see Appendix \ref{Appendix:Light_intensity_at_detector}), is proportional to the CL field,  $\left(I_2-I_1\right)\approx 2 \Re e \left[ E_R E_{CL} \right]$. Figure \ref{fig:CL_into_a_waveguide}c shows a calculation of the expected signal in each of the detectors per shot, and the systematic difference between them. The shot noise of the strong reference field dominates each individual detector (marked by their standard deviation $\sigma_{1,2}$). However, the noise is correlated, and thus vanishes when subtracting the signal of the two detectors (see Appendix \ref{Appendix:Intensity_at_balanced_PD}). The reference laser pulse acts as a heterodyning field for the weak CL signal, which enables the detection of the coherent-CL emission with shot-noise sensitivity \cite{Yuen_Chan_OL_1983_Noise_in_Hetero_Detection}. In other words, the suppressed noise from the reference field in the difference signal prevents its buildup in the absence of a CL emission. Therefore, the signal-to-noise ratio (SNR) is given by the intrinsic quantum fluctuations of the CL, combined with the quantum efficiency of the detectors \cite{Yuen_Chan_OL_1983_Noise_in_Hetero_Detection, abbas_local-oscillator_1983}. In addition, the reference pulse simply needs to overlap with the CL pulse, without a significant benefit for matching their duration. The next nonzero term of the noise (see Appendix \ref{Appendix:Intensity_at_balanced_PD}) can be smaller by orders of magnitude. Consequently, the limiting factor in an experiment would be the buildup of noise originating from imperfect or imbalanced quantum efficiency of the two detectors and from intensity fluctuations of the reference laser field \cite{abbas_local-oscillator_1983}.

Figure \ref{fig:CL_into_a_waveguide}d-f depict the amplitudes of the coupling and the field for various propagation lengths of the electron in the near field of the waveguide, alongside the emitted CL-field in Fig. \ref{fig:CL_into_a_waveguide}g-i. We note that for a long propagation path along the fiber, the degree of coherence (blue) can be wider than the CL coupling $g_\omega$ (red), leading to a lower SNR and to distortions in the temporal shape of the CL field, as shown in Fig. \ref{fig:CL_into_a_waveguide}i. In such a case, the CL-field emission is optimal for \textit{longer electron pulses} that match better to the bandwidth of the coupling. \\

In conclusion, we have shown that even though the emission rate of cathodoluminescence is independent of the temporal structure of the electron, information embedded on a modulated electron wave function can be retrieved by heterodyning with a reference field. Thus, the characterization of the CL emission can reach the shot-noise limit, allowing the detection of coherent CL-emission from a single electron. Conceptually, the proposed experimental system is a Mach-Zehnder-like interferometer, mixing the reference light from one arm with the CL-mediated optical coherence carried by the electron in the other arm. The CL reveals the extreme-nonlinear nature of coherence transfer by free electrons, which is capable of emitting a broadband spectrum with intricate correlations, including a degree of coherence for harmonics of the driving field. This concept holds potential for sensitive spectroscopy, as well as for the coherent manipulation of heterostructures and individual quantum systems with femtosecond optical resolution at the atomic scale.

\bibliography{Coherent_CL_v1}

\begin{thebibliography}{10}

\bibitem{Polman_Kociak_JGdA_NatMat_2019}
A.~Polman, M.~Kociak, and F.~J. Garc\'{i}a~de Abajo.
\newblock Electron-beam spectroscopy for nanophotonics.
\newblock {\em Nature Materials}, 18(11):1158--1171, November 2019.

\bibitem{Reimer_Kohl_TEM_book_2008}
L.~Reimer and H.~Kohl.
\newblock {\em Transmission {Electron} {Microscopy}: {Physics} of {Image}
  {Formation}}.
\newblock Springer {Series} in {Optical} {Sciences}. Springer-Verlag, New York,
  5 edition, 2008.

\bibitem{Barwick_PINEM_2009}
B.~Barwick, D.~J. Flannigan, and A.~H. Zewail.
\newblock Photon-induced near-field electron microscopy.
\newblock {\em Nature}, 462(7275):902--906, December 2009.

\bibitem{Park_NJP_PINEM_2010}
S.~T. Park, M.~Lin, and A.~H. Zewail.
\newblock Photon-induced near-field electron microscopy ({PINEM}): theoretical
  and experimental.
\newblock {\em New Journal of Physics}, 12(12):123028, December 2010.

\bibitem{JGdA_NanoLett_2010}
F.~J. Garc\'{i}a~de Abajo, A.~Asenjo-Garcia, and M.~Kociak.
\newblock Multiphoton {Absorption} and {Emission} by {Interaction} of {Swift}
  {Electrons} with {Evanescent} {Light} {Fields}.
\newblock {\em Nano Letters}, 10(5):1859--1863, May 2010.

\bibitem{H99}
A.~Howie.
\newblock Electrons and photons: exploiting the connection.
\newblock {\em Inst. Phys. Conf. Ser.}, 161:311--314, 1999.

\bibitem{EEGS_Javier_Kociak_NJP_2008}
F.~J. Garc\'{i}a~de Abajo and M.~Kociak.
\newblock Electron energy-gain spectroscopy.
\newblock {\em New Journal of Physics}, 10(7):073035, July 2008.

\bibitem{Piazza_EEGS_antenna_NatCom_2015}
L.~Piazza, T.~T.~A. Lummen, E.~Quiñonez, Y.~Murooka, B.~W. Reed, B.~Barwick,
  and F.~Carbone.
\newblock Simultaneous observation of the quantization and the interference
  pattern of a plasmonic near-field.
\newblock {\em Nature Communications}, 6:6407, March 2015.

\bibitem{DAS_Kociak_EEGS_Ultramic_2019}
P.~Das, J.~D. Blazit, M.~Tencé, L.~F. Zagonel, Y.~Auad, Y.~H. Lee, X.~Y. Ling,
  A.~Losquin, C.~Colliex, O.~Stéphan, F.~J. Garc\'{i}a~de Abajo, and
  M.~Kociak.
\newblock Stimulated electron energy loss and gain in an electron microscope
  without a pulsed electron gun.
\newblock {\em Ultramicroscopy}, 203:44--51, August 2019.

\bibitem{vanacore_transverse_PINEM_NatComm_2018}
G.~M. Vanacore, I.~Madan, G.~Berruto, K.~Wang, E.~Pomarico, R.~J. Lamb,
  D.~McGrouther, I.~Kaminer, B.~Barwick, F.~J. Garc\'{i}a~de Abajo, and
  F.~Carbone.
\newblock Attosecond coherent control of free-electron wave functions using
  semi-infinite light fields.
\newblock {\em Nature Communications}, 9(1):2694, July 2018.

\bibitem{vanacore_PINEM_vortex_NatMat_2019}
G.~M. Vanacore, G.~Berruto, I.~Madan, E.~Pomarico, P.~Biagioni, R.~J. Lamb,
  D.~McGrouther, O.~Reinhardt, I.~Kaminer, B.~Barwick, H.~Larocque, V.~Grillo,
  E.~Karimi, F.~J. Garc\'{i}a~de Abajo, and F.~Carbone.
\newblock Ultrafast generation and control of an electron vortex beam via
  chiral plasmonic near fields.
\newblock {\em Nature Materials}, 18(6):573--579, June 2019.

\bibitem{feist_transverse_PINEM_2020}
A.~Feist, S.~V. Yalunin, S.~Sch\"{a}fer, and C.~Ropers.
\newblock High-purity free-electron momentum states prepared by
  three-dimensional optical phase modulation.
\newblock {\em arXiv:2003.01938 [cond-mat, physics:physics, physics:quant-ph]},
  March 2020.
\newblock arXiv: 2003.01938.

\bibitem{Feist_Nature_2015}
A.~Feist, K.~E. Echternkamp, J.~Schauss, S.~V. Yalunin, S.~Schäfer, and
  C.~Ropers.
\newblock Quantum coherent optical phase modulation in an ultrafast
  transmission electron microscope.
\newblock {\em Nature}, 521(7551):200--203, May 2015.

\bibitem{Priebe_atto_SQUIRRELS_NatPhot_2017}
K.~E. Priebe, C.~Rathje, S.~V. Yalunin, T.~Hohage, A.~Feist, S.~Schäfer, and
  C.~Ropers.
\newblock Attosecond electron pulse trains and quantum state reconstruction in
  ultrafast transmission electron microscopy.
\newblock {\em Nature Photonics}, 11(12):793--797, December 2017.

\bibitem{Morimoto_atto_NatPhys_2018}
Y.~Morimoto and P.~Baum.
\newblock Diffraction and microscopy with attosecond electron pulse trains.
\newblock {\em Nature Physics}, 14(3):252, March 2018.

\bibitem{Kozak_Hommelhoff_atto_PRL_2018}
M.~Kozák, N.~Schönenberger, and P.~Hommelhoff.
\newblock Ponderomotive {Generation} and {Detection} of {Attosecond}
  {Free}-{Electron} {Pulse} {Trains}.
\newblock {\em Physical Review Letters}, 120(10):103203, March 2018.

\bibitem{Gover_and_Yariv_2LS_Rabi_after_PINEM_PRL2020}
A.~Gover and A.~Yariv.
\newblock Free-{Electron}–{Bound}-{Electron} {Resonant} {Interaction}.
\newblock {\em Physical Review Letters}, 124(6):064801, February 2020.

\bibitem{Coenen_Polman_APL_2011}
T.~Coenen, E.~J.~R. Vesseur, and A.~Polman.
\newblock Angle-resolved cathodoluminescence spectroscopy.
\newblock {\em Applied Physics Letters}, 99(14):143103, October 2011.
\newblock Publisher: American Institute of Physics.

\bibitem{Osorio_Polman_ACSPhot_2016}
C.~I. Osorio, T.~Coenen, B.~J.~M. Brenny, A.~Polman, and A.~F. Koenderink.
\newblock Angle-{Resolved} {Cathodoluminescence} {Imaging} {Polarimetry}.
\newblock {\em ACS Photonics}, 3(1):147--154, January 2016.

\bibitem{Chu_JGdA_Dark_plasmon2009}
M.-W. Chu, V.~Myroshnychenko, C.~H. Chen, J.-P. Deng, C.-Y. Mou, and F.~J.
  García~de Abajo.
\newblock Probing bright and dark surface-plasmon modes in individual and
  coupled noble metal nanoparticles using an electron beam.
\newblock {\em Nano Letters}, 9(1):399--404, 2009.
\newblock PMID: 19063614.

\bibitem{Javier_rev_mod_ph_2010}
F.~J. Garc\'{i}a~de Abajo.
\newblock Optical excitations in electron microscopy.
\newblock {\em Reviews of Modern Physics}, 82(1):209--275, February 2010.

\bibitem{OPticaDIgiulio}
V.~D. Giulio, M.~Kociak, and F.~J. Garc\'{i}a~de Abajo.
\newblock Probing quantum optical excitations with fast electrons.
\newblock {\em Optica}, 6(12):1524--1534, December 2019.

\bibitem{Mandel_and_Wolf}
L.~Mandel and E.~Wolf.
\newblock {\em Optical {Coherence} and {Quantum} {Optics}}.
\newblock Cambridge University Press, September 1995.
\newblock Google-Books-ID: dVYhAwAAQBAJ.

\bibitem{Glauber_coherent_1963}
R.~J. Glauber.
\newblock Coherent and {Incoherent} {States} of the {Radiation} {Field}.
\newblock {\em Physical Review}, 131(6):2766--2788, September 1963.

\bibitem{Feist_UTEM_Ultramicroscopy_2017}
A.~Feist, N.~Bach, N.~Rubiano~da Silva, T.~Danz, M.~Möller, K.~E. Priebe,
  T.~Domröse, J.~G. Gatzmann, S.~Rost, J.~Schauss, S.~Strauch, R.~Bormann,
  M.~Sivis, S.~Schäfer, and C.~Ropers.
\newblock Ultrafast transmission electron microscopy using a laser-driven field
  emitter: {Femtosecond} resolution with a high coherence electron beam.
\newblock {\em Ultramicroscopy}, 176:63--73, May 2017.

\bibitem{Kfir_e_ph_entanglement_PRL2019}
O.~Kfir.
\newblock Entanglements of {Electrons} and {Cavity} {Photons} in the
  {Strong}-{Coupling} {Regime}.
\newblock {\em Physical Review Letters}, 123(10):103602, September 2019.

\bibitem{Bendana_Polman_JGdA_CL_fiber_NL_2010}
X.~Benda\~{n}a, A.~Polman, and F.~J. Garc\'{i}a~de Abajo.
\newblock Single-{Photon} {Generation} by {Electron} {Beams}.
\newblock {\em Nano Letters}, 11(12):5099--5103, December 2011.

\bibitem{Yuen_Chan_OL_1983_Noise_in_Hetero_Detection}
H.~P. Yuen and V.~W.~S. Chan.
\newblock Noise in homodyne and heterodyne detection.
\newblock {\em Optics Letters}, 8(3):177--179, March 1983.
\newblock Publisher: Optical Society of America.

\bibitem{abbas_local-oscillator_1983}
G.~L. Abbas, V.~W.~S. Chan, and T.~K. Yee.
\newblock Local-oscillator excess-noise suppression for homodyne and heterodyne
  detection.
\newblock {\em Optics Letters}, 8(8):419--421, August 1983.
\newblock Publisher: Optical Society of America.

\bibitem{E_diff_vac_fluct_DiGiulio_Abajo_NJP2020}
F.~J. Garc\'{i}a~de Abajo and V.~Di~Giulio.
\newblock Electron {Diffraction} by {Vacuum} {Fluctuations}.
\newblock {\em New Journal of Physics}, 2020.

\bibitem{dung_three-dimensional_quantization_1998}
H.~T. Dung, L.~Knöll, and D.-G. Welsch.
\newblock Three-dimensional quantization of the electromagnetic field in
  dispersive and absorbing inhomogeneous dielectrics.
\newblock {\em Physical Review A}, 57(5):3931--3942, May 1998.

\bibitem{digiulio2020freeelectron}
V.~Di~Giulio and F.~J. Garc\'{i}a~de Abajo.
\newblock Free-electron shaping using quantum light, 2020.

\end{thebibliography}

\newpage

\appendix
\section{Detailed derivations \label{Appendix:A}}

\subsection{Evolution operator\label{Appendix:Scatt_operator_derivation}}

We consider an electron beam that is well described by a wave-packet localized in momentum space around a wave vector $\kb_0$. If we further assume that the photon energies involved are small compared with the electron relativistic energy $E_0=c\sqrt{\hbar^2 m^2+\hbar^2k_0^2}$ and working in a gauge with zero scalar potential, the Hamiltonian describing the evolution of the system is \cite{OPticaDIgiulio}
\begin{align}
\hat{\mathcal{H}}_0&=\sum_{\kb} \left[E_0+ \hbar \vb \cdot (\kb - \kb_0)\right]c^\dagger_\kb c_\kb + \mathcal{H}_0^{\rm F},\nonumber \\
\hat{\mathcal{H}}_{\rm int}&=\frac{-1}{c}\sum_{\alpha}\int d^3\rb \hat{J}_\alpha (\rb) \hat{A}_\alpha (\rb)\nonumber 
,\end{align}
where we have defined the current operator $\hat{J}_\alpha(\rb)=(-e v_\alpha/V)\sum_{\kb,\qb} \ee^{\ii \qb \cdot \rb}c^\dagger_\kb c_{\kb+\qb}$ using $c_{\kb}$ Fermionic ladder operators and the relativistic electron group velocity $\vb=\hbar c^2 \kb_0/E_0$. Therefore, the scattering operator governing the evolution of the system from time $t=-\infty$ to $t=\infty$ is given by Di Giulio et al., (in preparation).
\begin{align}
\hat{\mathcal{S}}(\infty,-\infty)=\ee^{\ii \hat{\chi}(\infty,-\infty)}\hat{\mathcal{U}}(\infty,-\infty)\nonumber
,\end{align}
where
\begin{align}
\hat{\mathcal{U}}(\infty,-\infty)&=\ee^{\int_0^\infty d\omega g_\omega(\hat{b}_\omega\hat{a}_\omega^\dagger  -\hat{b}_\omega^\dagger \hat{a}_\omega)}\nonumber\\
\hat{b}_\omega &=\sum_{k_z}c^\dagger_{k_z} c_{k_z+\omega/v},\nonumber \\
\hat{a}_\omega &= \ii \frac{2e\omega}{\hbar g_\omega }\int_{-\infty}^\infty dz ~\ee^{-\ii \omega z / v}\int d^3 \rb' \sqrt{\hbar {\rm Im}\{\epsilon(\rb',\omega)\}}\sum_{i} G_{z,i}(\Rb_0,z,\rb',\omega)\hat{f}_i(\rb',\omega)  \nonumber
.\end{align}
We have further assumed an incident focused electron with a transversal component of the wave function satisfying $|\psi_i(\Rb)|^2 
\approx \delta(\Rb-\Rb_0)$, and $g_\omega = \sqrt{\Gamma^{\rm EELS}(\omega)}$ with $\Gamma^{\rm EELS}(\omega)$ being the real-valued electron-energy-loss probability \cite{Javier_rev_mod_ph_2010}. The operator $ \hat{\chi}(\infty,-\infty)$ does not need to be specified because it only acts on the electron degrees of freedom \cite{E_diff_vac_fluct_DiGiulio_Abajo_NJP2020} and it is not of interest for this study. Additionally, the previous operators satisfy the following commutation relations:
\begin{align}
[\hat{a}_\omega , \hat{a}^\dagger_{\omega '}]&=\delta(\omega - \omega'),\nonumber\\
[\hat{b}_\omega,\hat{b}^\dagger_{\omega'}]&=0,\nonumber\\
\left[f_i(\rb,\omega),f_j^\dagger(\rb',\omega')\right]&=\delta(\rb-\rb')\delta(\omega-\omega')\delta_{i,j},\nonumber
\end{align}
which are used in the derivation of the following results. The commutator $[\hat{a}_\omega , \hat{a}^\dagger_{\omega '}]$ is insensitive to the phase of the optical mode, which allows us to define a non-negative $g_\omega$. In the main text, there is a preferred selection of the optical-mode phase, and hence $g_\omega$ can be complex. In the example of waveguide modes, as in Fig. \ref{fig:CL_into_a_waveguide}h, the side lobes are a direct result of a sign change of $g_\omega$ within the bandwidth of the coherent emission. 

\subsection{Initial joint electron-sample state}
If we consider the incoming electron being in a superposition of energy states and the sample starting from the ground state (i.e., zero photons), the electron-sample initial state in the interaction picture can be written as 
\begin{align}
|\psi(-\infty)\rangle = \sum_{k_z} \alpha_{k_z} |k_z,0\rangle,     \nonumber
\end{align}
which corresponds to Eq. \eqref{eq:e_ladder_state} expressed in a basis of electron energies.

\subsection{Mean number of excitations after interaction \label{Appendix:CL_photon_number}}
By evolving the initial state to the final state $|\psi(\infty)\rangle = \hat{S}(\infty,-\infty)|\psi(-\infty)\rangle$, we can calculate the average number of excitations, which is independent of the electron wave function,
\begin{align}
\langle \hat{a}_\omega^\dagger \hat{a}_\omega \rangle &=\langle \psi(\infty )| \hat{a}_\omega^\dagger \hat{a}_\omega |\psi(\infty)\rangle\nonumber \\
&= g^2_\omega \nonumber
.\end{align}

\subsection{Mean electric field after interaction \label{Appendix:Mean_electric_field_Valerio}}
The quantum average (or the expectation value) of the positive-energy electric field operator, defined as \cite{dung_three-dimensional_quantization_1998}
\begin{align}
\hat{E}_i^{(+)}(\rb,\omega,t=0)=-4 \pi \ii \omega^2 \int d^3 \rb' \sqrt{\hbar {\rm Im}\{\epsilon(\rb',\omega)\}}\sum_j G_{i,j}(\rb,\rb',\omega) \hat{f}_j(\rb',\omega ), \label{eq:cloperator}
\end{align}
which gives
\begin{align}
\langle \hat{E}_i^{(+)}(\rb,\omega,t=0) \rangle =8\pi e \omega M_i(\rb, \omega) \int_{-\infty}^\infty dz \ee^{-\ii\omega z /v}|\psi(z,t=-\infty) |^2, \nonumber  
\end{align}
where we have defined $M_i(\rb,\omega)=\int_{-\infty}^{\infty} dz \ee^{\ii \omega z /v} {\rm Im}\{G_{i,z}(\rb,\Rb_0,z,\omega)\}$ and introduced the incoming electron wave function in the interaction picture $\psi(z,-\infty)=\sum_{k_z}\alpha_{k_z}\ee^{\ii k_z z}/\sqrt{L}$ with $L$ denoting the quantization length. Interestingly, this expression can be recast by using the relation
\begin{align}
    \langle \hat{a}_\omega \rangle= g_\omega \int_{-\infty}^\infty dz \ee^{-\ii\omega z /v}|\psi(z,t=-\infty) |^2. \nonumber
\end{align}
This result is equivalent to Eq. \eqref{eq:E_g_nw_FT_psi_sq} in the main text, and thus,
\begin{align}
 \langle \hat{E}_i^{(+)}(\rb,\omega,t=0) \rangle = 8\pi e \omega  \langle \hat{a}_\omega \rangle M_i(\rb, \omega)/g_\omega. \label{eq:eplus}
.\end{align}

\subsection{Degree of coherence \label{Appendix:DOC_Continuum_states_Valerio}}
In order to quantify the coherence transmitted from the modulated electron to the sample excitations, we define the degree of coherence per unit frequency as
\begin{align}
{\rm DOC}(\omega)= \frac{\langle \hat{E}_i^{(-)}(\rb,\omega,t=0)\rangle \langle \hat{E}_i^{(+)}(\rb,\omega,t=0) \rangle}{\langle \hat{E}_i^{(-)}(\rb,\omega,t=0) \hat{E}_i^{(+)}(\rb,\omega,t=0) \rangle}.\nonumber
\end{align}
This quantity can be calculated from Eq.\ (\ref{eq:eplus}) together with the expression
\begin{align}
\langle \hat{E}_i^{(-)}(\rb,\omega,t=0) \hat{E}_i^{(+)}(\rb,\omega,t=0) \rangle =64 \pi^2e^2 \omega^2 |M_i(\rb,\omega)|^2  \nonumber
\end{align}
for the denominator. We find 
\begin{align}
    {\rm DOC}(\omega) &= \frac{\langle \hat{a}_\omega^\dagger \rangle \langle \hat{a}_\omega\rangle}{\langle\hat{a}_\omega^\dagger  \hat{a}_\omega\rangle }\nonumber\\
    &= \left| \int_{-\infty}^\infty dz \ee^{-\ii \omega z /v}|\psi(z,t=-\infty) |^2 \right|^2,   \nonumber
\end{align}
as in Eq. \eqref{eq:defining_DOC} in the main text.
Incidentally, we note that the spatial dependence of the degree of coherence cancels, thus making it a spectral property. 

\subsection{Degree of coherence for a PINEM-modulated electron \label{Appendix:DOC_PINEM_explicit}}
It is interesting to study the degree of coherence for an electron that is modulated by free propagation after PINEM interaction with a laser field. Here we assume a PINEM laser frequency $\omega_0$ and a PINEM coupling parameter $\beta$\cite{Polman_Kociak_JGdA_NatMat_2019}. In particular, beyond the PINEM interaction region, the electron wave function in the Schr\"{o}dinger picture reduces to 
\begin{align}
    \psi^{(\rm S)}(z,t)= \frac{\phi(z- vt)}{\sqrt{L}}\ee^{-\ii \varepsilon_0 t } \sum_{\ell=-\infty}^\infty J_\ell (2|\beta|)\ee^{-\ii \ell \omega_0  t + \ii k_\ell z +\ii \ell{\rm arg\{-\beta\}}},\nonumber
\end{align}
where $\varepsilon_0=E_0/\hbar$, $k_\ell = \sqrt{E_\ell^2/c^2-m^2c^2}/\hbar$ with $E_\ell=E_0+\hbar \omega_0 \ell$, and $L$ is the quantization length (see Ref. \cite{digiulio2020freeelectron}). Therefore, the Fourier components of the wave function are given by
\begin{align}
\alpha^{\rm (S)}_k(t)&=\int_{-\infty}^{\infty}dz \psi^{(\rm S)}(z,t)\frac{\ee^{-\ii k z}}{\sqrt{L}}\nonumber\\
&=\frac{\ee^{-\ii \varepsilon_0 t}}{L}\sum_\ell J_\ell(2|\beta|)\ee^{\ii (k_0-k-2\pi \ell^2/z_{\rm T})vt}\ee^{\ii \ell{\rm arg\{-\beta\}}}\int_{-\infty}^\infty d\tilde{z} \ee^{\ii (k_0 +\ell \omega_0/v -k)\tilde{z}}\phi(\tilde{z})\nonumber,
\end{align}
where we have expanded the wave vector up to second order in $\ell$ and we have used the fact that $\abs{z-vt} \ll z_{\rm T}=4\pi m v^3\gamma^3/\hbar\omega_0^2$ within the interaction region. If we consider an electron pulse of infinite duration, we can approximate $\phi(\tilde{z})\sim 1$, which yields
\begin{align}
\alpha^{\rm (S)}_k(t)=\ee^{-\ii\varepsilon_0 t}\sum_\ell J_\ell (2|\beta|)\ee^{\ii (k_0-k-2\pi \ell^2/z_{\rm T})vt}\ee^{\ii \ell{\rm arg\{-\beta\}}}\delta_{k_0+\ell \omega_0/v,k},\nonumber
\end{align}
where we have used the relation $\int dz \ee^{\ii(k-k')z}=L\delta_{k,k'}$. We can now move to the interaction picture by multiplying $\alpha^{(\rm S)}_k(t)$ by the factor $\ee^{\ii [\varepsilon_0 + v(k-k_0)]t}$. Additionally, by going back to real space, we obtain the electron wave function in the interaction picture
\begin{align}
    \psi(z,t=-\infty)=\sum_\ell J_\ell(2|\beta|)\ee^{- 2\pi \ii \ell^2 d /z_{\rm T}}\ee^{\ii \ell{\rm arg\{-\beta\}}}\frac{\ee^{\ii(k_0+\ell\omega_0/v)z}}{\sqrt{L}},\nonumber
\end{align}
where we have substituted $vt$ by the propagation distance from the PINEM interaction region $d$. 
We are now ready to calculate the Fourier transform of the electron density, which reads
\begin{align}
 \int_{-\infty}^\infty dz \ee^{-\ii \omega z /v}|\psi(z,t=-\infty)|^2 = \sum_{\ell,\ell'}J_\ell(2|\beta|)J_{\ell'}(2|\beta|)\ee^{\ii (\ell-\ell'){\rm arg\{-\beta\}}}\ee^{-2\pi \ii (\ell^2 -\ell'^2)d/z_{\rm T}}\delta_{\omega_0(\ell-\ell'),\omega}. \label{eq:densft}
\end{align}
From Eq. \eqref{eq:densft}, we immediately see that the integral vanishes unless $\omega=n\omega_0$, where $n$ is an integer. Thus, the frequency of the emitted light that shows coherence with respect to the PINEM-driving laser is a multiple of the PINEM laser frequency. With this assumption, we obtain
\begin{align}
     \int_{-\infty}^\infty dz \ee^{-\ii \omega z /v}|\psi(z,t=-\infty)|^2 = \ee^{\ii n{\rm arg\{-\beta\}}} \ee^{2 \pi \ii n^2 d/z_{\rm T}}\sum_\ell J_\ell(2|\beta|)J_{\ell-n}(2|\beta|)\ee^{-4\pi \ii \ell n d/z_{\rm T}}.\label{eq:densft2}
\end{align}
Clearly, all the quantities proportional to the Fourier transform in Eq. (\ref{eq:densft2}) are calculated for a given harmonic order $n=\omega/\omega_0$. 

\subsection{On the calculation of quantum averages of time-dependent field operators \label{Appendix:Light_intensity_at_detector}}
According to Glauber's prescription \cite{Glauber_coherent_1963}, in order to calculate averages of time-dependent operators to then compute measurable quantities (e.g., light intensities and correlation functions),  the operators have to be understood in the Heisenberg picture. For instance, the time-varying light intensity at a given point in space $\rb$ for polarization $\alpha$ has to be calculated as
\begin{align}
I_\alpha(\rb,t)=\frac{c}{2\pi}\langle \psi_H| E_\alpha^{H,(-)}(\rb,t)E_\alpha^{H,(+)}(\rb,t)|\psi_H\rangle\nonumber
,\end{align} 
where the subscript $H$ stands for Heisenberg picture. Now, by using an adiabatic switching of the interaction, which provides the connection between the interaction and Heisenberg pictures $\ee^{-\ii \mathcal{H} t}=\ee^{-\ii \mathcal{H}_0 t}\hat{S}(t,-\infty)$, we can rewrite the average in terms of scattering operators
 \begin{align}
I_\alpha(\rb,t)=\frac{c}{2\pi}\langle \psi(-\infty)|\hat{S}^\dagger(t,-\infty) \hat{E}_\alpha^{(-)}(\rb,t)\hat{E}_\alpha^{(+)}(\rb,t)\hat{S}(t,-\infty)|\psi(-\infty)\rangle.\nonumber
\end{align} 

From the previous expression, it is clear that the scattering matrix $\hat{S}(\infty,-\infty)$ never appears unless we calculate the intensity at $t=\infty$, which is not the quantity in which one is usually interested. However, a simplification that can be used in this study consists of considering  the time-dependent field observables at large times, thus neglecting the few-femtosecond transient period in which the electron is still interacting with the sample and producing nonzero quantum averages of the electric field and light intensity. By doing so, we can extend the final time in the scattering operator to infinity, which leads to
\begin{align}
I_\alpha(\rb,t)\approx& \frac{c}{2\pi}\langle \psi(-\infty)|\hat{S}^\dagger(\infty,-\infty) \hat{E}_\alpha^{(-)}(\rb,t)\hat{E}_\alpha^{(+)}(\rb,t)\hat{S}(\infty,-\infty)|\psi(-\infty)\rangle
\nonumber \\
=& \frac{c}{2\pi} \langle \hat{E}_\alpha^{(-)}(\rb,t)\hat{E}_\alpha^{(+)}(\rb,t) \rangle.\label{eq:approxintensity}
\end{align}
We finally remark that all the time-dependent quantities in this work imply this assumption.

\subsection{Intensity and noise in a balanced detection experiment for finite electron and laser pulses
\label{Appendix:Intensity_at_balanced_PD}}
In this section, we compute the signal and noise in a balanced detection experiment. We assume that a replica of the the PINEM-driving laser field is the reference field, and that the weak cathodoluminescence emanating from interaction of the PINEM-modulated electron with a sample placed downstream is the signal. These two fields are mixed in a symmetric beam splitter, as depicted in Fig. \ref{fig:TEM_MachZehnder}. The beam splitter is characterized by reflection and transmission coefficients $R$ and $T$, which we assume to be independent of the optical frequency or polarization. Subsequently, the total signal in collected by two ideal detectors, labelled $\rm D1$ and $\rm D2$, respectively. We write the reference electric field operator as
\begin{align}
    \hat{E}_\alpha^{\rm R}(\rb,t)=\int_0^\infty \frac{d\omega}{2\pi} \mathcal{E}_\alpha^{\rm R}(\rb,\omega)\hat{a}_\omega^{\rm R} \ee^{-\ii \omega t} +{\rm h.c.} \nonumber,
\end{align}
where the electric field operator connected to the cathodoluminescence emission, $\hat{E}_\alpha^{\rm CL}(\rb,t)$, is given by the time dependent analog of Eq. \eqref{eq:cloperator}. We remark that the reference field operators $\hat{a}_\omega^{\rm R}$ and $\hat{a}_\omega^{{\rm R}\dagger}$ are assumed to also satisfy the commutation relation $[\hat{a}_\omega^{\rm R},\hat{a}^{{\rm R}\dagger}_{\omega'}]=\delta(\omega-\omega')$. The intensity operator at detector D1 is
\begin{align}
\hat{I}_\alpha^{\rm D1}(\rb,t)=& \frac{c}{2\pi}\left(T^* \hat{E}^{(-),{\rm CL}}_\alpha(\rb,t)+R^*\hat{E}^{(-),{\rm R}}_\alpha(\rb,t) \right)\left(T \hat{E}^{(+),{\rm CL}}_\alpha(\rb,t)+R\hat{E}^{(+),{\rm R}}_\alpha(\rb,t) \right).\nonumber
\end{align}
Intuitively, the operator $\hat{I}^{\rm D2}_\alpha$ can be obtained from $\hat{I}^{\rm D1}_\alpha$ by exchanging the $T$ and $R$ coefficients. The fluence (optical energy per unit area) impinging on each detector is 
\begin{align}
    \hat{F}_\alpha^{\rm D1/D2}(\rb)=\int_{-\infty}^\infty dt \hat{I}_\alpha^{\rm D1/D2}(\rb,t). \nonumber 
\end{align}
The difference between the two detectors, which is our signal, is given by
\begin{align}
     \mathcal{S}&=\langle \hat{F}_\alpha^{\rm D1}(\rb)\rangle-\langle \hat{F}_\alpha^{\rm D2}(\rb)\rangle \\
     &=\frac{2 e c}{\pi}{\rm Re}\left\{(R^*T-RT^*)\int_0^\infty d\omega  \mathcal{E}_\alpha^{\rm R *}(\rb,\omega) \alpha^{\rm R *}(\omega) M_\alpha(\rb,\omega)\left(\int_{-\infty}^{\infty} dz\ee^{-\ii \omega z/v}|\psi(z,t=-\infty)|^2\right) \right\}.\nonumber 
\end{align}
We redefined here the average $\langle \cdot \rangle$ symbol such that it includes a continuous mode coherent state $|\{\alpha^{\rm R}\}\rangle$, where $\hat{a}_\omega^{\rm R}|\{\alpha^{\rm R}\}\rangle=\alpha^{\rm R}(\omega)|\{\alpha^{\rm R}\}\rangle$. $\alpha^{\rm R}(\omega)$ is the frequency profile of the reference light pulse. One can choose, for example, Fresnel coefficients such as $R=1/\sqrt{2}$ and $T=\ii/\sqrt{2}$ to retrieve the heterodyne detection signal.  

The noise of such measurement can be calculated in a similar fashion. The square of the noise is defined by the variance as
\begin{align}
\mathcal{N}^2=\left\langle \left[\hat{F}_\alpha^{\rm D1}(\rb)-\hat{F}_\alpha^{\rm D2}(\rb)\right]^2\right\rangle-\left\langle \hat{F}_\alpha^{\rm D1}(\rb)-\hat{F}_\alpha^{\rm D2}(\rb)\right\rangle^2\label{eq:noise}.
\end{align}
Since we consider a strong reference laser field, $\int_{-\infty}^\infty |\alpha^{\rm R}|^2 d\omega \gg 1$, $\alpha^{\rm R}$ dominates Eq. \eqref{eq:noise}. For the choise $|R|=|T|$, the terms for the noise variance scaling as $|\alpha^{\rm R}|^4$ and $|\alpha^{\rm R}|^3$ vanish. Thus, the next leading order is substantially smaller and comprises combinations in which $\hat{E}^{\rm R}_\alpha$ and $\hat{E}^{\rm CL}$ appear twice each:
\begin{align}
    \mathcal{N}^2=&2\left(\frac{c}{4\pi^2}\right)^2\left(R^*T-RT^*\right)^2\int_0^\infty d\omega \int_0^\infty d\omega'\nonumber \\
    \times& {\rm Re} \left\{ \mathcal{E}_\alpha^{\rm R }(\rb,\omega) \alpha^{\rm R }(\omega) \mathcal{E}_\alpha^{\rm R }(\rb,\omega') \alpha^{\rm R }(\omega') \left[\langle \hat{E}_\alpha^{(-),\rm CL}(\rb,\omega) \hat{E}_\alpha^{(-),\rm CL}(\rb,\omega')\rangle-\langle\hat{E}_\alpha^{(-),\rm CL}(\rb,\omega) \rangle \langle\hat{E}_\alpha^{(-),\rm CL}(\rb,\omega') \rangle \right]\right. \nonumber \\
    &+ \mathcal{E}_\alpha^{\rm R }(\rb,\omega) \alpha^{\rm R }(\omega) \mathcal{E}_\alpha^{\rm R *}(\rb,\omega') \alpha^{\rm R *}(\omega')\langle \hat{E}_\alpha^{(-), \rm CL}(\rb,\omega) \rangle \langle \hat{E}_\alpha^{(+), \rm CL}(\rb,\omega') \rangle\nonumber \\
    &-\frac{1}{2}\mathcal{E}_\alpha^{\rm R }(\rb,\omega) \mathcal{E}_\alpha^{\rm R * }(\rb,\omega') \left[\delta(\omega - \omega')+\alpha^{\rm R }(\omega)\alpha^{\rm R *}(\omega')\right] \langle \hat{E}_\alpha^{(-), \rm CL}(\rb,\omega)  \hat{E}_\alpha^{(+), \rm CL}(\rb,\omega') \rangle  \nonumber\\
    &-\frac{1}{2}\alpha^{\rm R *}(\omega)\alpha^{\rm R }(\omega')\mathcal{E}_\alpha^{\rm R *}(\rb,\omega)\mathcal{E}_\alpha^{\rm R  }(\rb,\omega')\langle  \hat{E}_\alpha^{(+), \rm CL}(\rb,\omega) \hat{E}_\alpha^{(-), \rm CL}(\rb,\omega')\rangle \Big\}\nonumber,
\end{align}
where $t=0$ is implicitly understood in the CL operators. The contribution of this leading term to the noise is smaller by the ratio of the reference and CL fields, which could be many orders of magnitude smaller than the shot noise of the reference on each detector. Thus, this result implies a theoretical limit to the noise floor, independent of the intrinsic noise of the detectors. We conclude that the corresponding ideal signal-to-noise ratio in this system is ${\rm SNR}=\mathcal{S}/\mathcal{N}$.

\subsection{Explicit derivation of Eq. \eqref{eq:E_g_nw_FT_psi_sq} \label{Appendix:sum_cj_cj_prime}}
One can separate the initial electron-photon state in Eq. \eqref{eq:e_ladder_state} as the product of the electron part $\ket{\psi_e}$ and the vacuum of the radiation, $\ket{\psi_{in}}=\ket{\psi_e}\otimes\ket{0}$, where $\ket{\psi_e}=\sum_j c_j \ket{E_j}$. The ladder coefficients can be derived through simple algebra,
\begin{align*}
     \expect{E_\ell | \psi_{e}} 
     &= \round{ \sum_j c_j \underbrace{\expect{E_\ell | E_j}}_{\delta_{\ell,j}}} \\
     c_j 
     &=\expect{E_j | \psi_{e}}. 
\end{align*}
 We can now write the sum in Eq. \eqref{eq:a_vs_omega} as 
\begin{align*}
\sum_j c_j^* c_{j+n} 
&= \sum_j \expect{\psi_e | E_j} \expect{ E_{j+n} | \psi_e}\\
&= \expect{\psi_e \left|  \sum_j \ket{E_j}\bra{E_{j+n}} \right| \psi_e}   \\
&= \expect{\psi_e \left| e^{\ii n\omega_0 t}  \sum_j \ket{E_j}\bra{E_{j}} \right| \psi_e}   \\
&= \expect{\psi_e \left| e^{\ii n\omega_0 t} \right| \psi_e}  
.\end{align*}
Here, we assume that the energy states are identical, aside from their energy difference, and that $\ket{\psi_e}$ is fully spanned by the discrete and complete set of $\ket{E_j}$ states, that is, 
\begin{align*}
    \ket{E_{j+n}} &= e^{-\ii n\omega_0 t}\ket{E_j}   ,\\
    \sum_j\ket{E_j}\bra{E_j} &= \mathcal{I}
.\end{align*} 

In the sample region, we can approximate the electron dispersion as linear, and thus, the wave function can be written as a function of time, $\psi(t)$, relying on the relation $z - v_g t = constant $. Using this temporal electron wave function $\psi(t)$, one finds
\begin{align}
    \sum_j c_j^* c_{j+n} 
&= \int  \left(\psi(t)\right)^* e^{\ii in\omega_0 t} \psi(t) dt \nonumber \\
&= \int \abs{\psi(t)}^2 e^{\ii n\omega_0 t} dt = \mathcal{FT}\left[ \abs{\psi(t)}^2 \right]_{(n\omega_0)} \label{eq:cj_cj_prime_is_FT}
.\end{align}

\subsection{Generalization to strong electron-photon coupling \label{Appendix:strong_e_ph_coupling}}
Starting from the scattering operator in Eq. \eqref{eq:S_operator}, we are interested in the expectation value for a specific frequency $\Omega$, that is, $\left<   \hat{a}_\Omega  \right> $, as well as higher order terms. For this purpose, it is convenient to use the commutation relation  
\begin{equation*}  
\left[\hat{a}_\Omega, \hat{S} \right] = \left[\hat{a}_\Omega,e^{ \int_{{ \rm 0}}^{\infty} d\omega  \left( g_\omega \hat{b}_\omega \hat{a}_\omega^\dagger  -  g_\omega^*\hat{b}_\omega^\dagger \hat{a}_\omega  \right)  } \right]
.\end{equation*}
Since $\left[\hat{b}_\omega, \hat{b}_{\omega'}^\dagger\right]=\left[\hat{b}_\omega, \hat{S}\right]=0$ and 
\begin{equation*}
\left[\hat{a}_\Omega, \left( g_\omega \hat{b}_\omega \hat{a}_\omega^\dagger  -g_\omega^* \hat{b}_\omega^\dagger \hat{a}_\omega  \right)    \right] 
= \left[\hat{a}_\Omega,  g_\omega \hat{b}_\omega \hat{a}_\omega^\dagger   \right]
= g_\omega \hat{b}_\omega \delta(\omega-\Omega)
,\end{equation*}
one can use the conditional identity 
\begin{equation*}
\left[A, B\right] = c \quad \Rightarrow \quad \left[A, e^B\right]=ce^B  
,\end{equation*}
where $c$ is a c-number operator write
\begin{equation}\label{Commutation_a_S}
\left[\hat{a}_\Omega,\hat{S} \right] = g_\Omega  \hat{b}_\Omega \hat{S}
.\end{equation} 
The expectation value for $\left<\hat{a}_\Omega\right>$ on the final state, $\left|\psi_f\right>=\hat{S} \left| \psi_{initial} \right> = \hat{S}\sum_j c_j  \left| E_j,0 \right>$, is
\begin{align*}
\left< \hat{a}_\Omega \right>    &=     \sum_{j,j'}c_j^*c_{j'}\left<E_j,0\right| \hat{S}^\dagger \hat{a}_\Omega \hat{S} \left|E_{j'},0\right>.
\end{align*}
The commutation relation above allows us to write $\hat{S}^\dagger \hat{a}_\Omega \hat{S}=g_\Omega \hat{b}_\Omega + \hat{a}_\Omega$, so
\begin{align*}
\left< \hat{a}_\Omega \right> &=   \sum_{j,j'}c_j^*c_{j'}\left<E_j,0\right| g_\Omega \hat{b}_\Omega + \hat{a}_\Omega  \left|E_{j'},0\right>\\
&=  g_\Omega \sum_{j,j'}c_j^*c_{j'} \underbrace{        \left<E_j,0|E_{j'} - \hbar\Omega,0\right>       }_{\delta_{j,j'-n}, \text{ for }n=\Omega/\omega_0}\\
&=   g_\Omega  \sum_{j}c_j^*c_{j+n}\\
&\overset{\text{(eq.\,\eqref{eq:cj_cj_prime_is_FT})}}{=}   g_\Omega  \mathcal{FT}\left[\left|\psi(t)\right|^2\right]_{\left(\Omega=n \omega_0\right)}
,\end{align*}
as in Eq. \eqref{eq:E_g_nw_FT_psi_sq} in the main text for weak coupling. We have simplified this expression mostly by using $\hat{a}_\Omega  \left|E_{j'},0\right>=0$.

\subsection{Higher-order correlations\label{Appendix:higher_order_correlations}}
The moment of order $N$ for the quantum correlations of the CL is 
\begin{align}
\left<(\Delta \hat{a}_{{ \rm \Omega}})^N\right>&=\left<\left(\hat{a}_{{  \rm \Omega}}-\right<\hat{a}_{{  \rm \Omega}}\left>\right)^N\right> \nonumber \\
  &=    \sum_k\left<\binom{N}{k}\quad \hat{a}_{{  \rm \Omega}}^k \quad \left<-\hat{a}_{{  \rm \Omega}}\right>^{N-k}\right>  \nonumber \\
  &=    \sum_k\binom{N}{k}\left(-1\right)^{N-k}\left<\hat{a}_{{  \rm \Omega}}\right>^{N-k}\left< \hat{a}_{{  \rm \Omega}}^k  \right> \label{binom_high_order_a},
\end{align}
where $\binom{N}{k}$ denotes Newton's binomial coefficients. The operator $ \left< \hat{a}^k_\Omega \right>$ can be simplified using the unitarity of $\hat{S}$, that is,  $\hat{S}^\dagger \hat{S}=\hat{I}$, and substituting $\hat{S}^\dagger \hat{a}_\Omega^k \hat{S}=\left(\hat{S}^\dagger \hat{a}_\Omega \hat{S}\right)^k  $. We find
\begin{align*}
\left< \hat{a}_\Omega ^k\right>    
&=     \sum_{j,j'}c_j^*c_{j'}\left<E_j,0\right| \hat{S}^\dagger \hat{a}_\Omega^k \hat{S} \left|E_{j'},0\right>\\
&=     \sum_{j,j'}c_j^*c_{j'}\left<E_j,0\right| \left(\hat{S}^\dagger \hat{a}_\Omega \hat{S} \right)^k \left|E_{j'},0\right>\\
&\overset{(*)}{=}     \sum_{j,j'}c_j^*c_{j'}\left<E_j,0\right| \left(     
g_\Omega \hat{b}_\Omega + \hat{a}_\Omega
\right)^k \left|E_{j'},0\right>\\
&=    g_\Omega^k \sum_{j,j'}c_j^*c_{j'}  \underbrace{\left<E_j,0\right|    
 \hat{b}_\Omega^k \left|E_{j'},0\right>}_{\delta_{j',j+k\Omega/\omega_0}}\\
&= g_\Omega^k \sum_j c_j^* c_{j+k\frac{\Omega}{\omega_0}}
.\end{align*} 
The equation marked with $(*)$ uses the commutation in Eq. \eqref{Commutation_a_S}. 
We can now substitute $\left< \hat{a}_\Omega^k \right>$ and $\left< \hat{a}_\Omega \right>$ in Eq. \eqref{binom_high_order_a} to write

\begin{align*}
\left<(\Delta \hat{a}_{{  \rm \Omega}})^N\right>
  &=    \sum_k\binom{N}{k}\left(-1\right)^{N-k}\left<\hat{a}_{{  \rm \Omega}}\right>^{N-k}\left< \hat{a}_{{  \rm \Omega}}^k  \right> \\
&=    \sum_k\binom{N}{k}\left(-1\right)^{N-k}\left( g_\Omega \sum_j c_j^* c_{j+\frac{\Omega}{\omega_0}}\right)^{N-k}\left( g_\Omega^k \sum_j c_j^* c_{j+k\frac{\Omega}{\omega_0}}  \right)\\
&=   g_\Omega^N \sum_k\binom{N}{k}\left(-1\right)^{N-k}\left(  \mathcal{FT}\left[\left|\psi(t)\right|^2\right]_{\left(\Omega\right)}  \right)^{N-k}\left(  \mathcal{FT}\left[\left|\psi(t)\right|^2\right]_{\left(k\Omega\right)}  \right)
.\end{align*}

\end{document}